# THE EFFECTS OF INTERPLAY BETWEEN THE ROTATION AND SHOALING FOR A SOLITARY WAVE ON VARIABLE TOPOGRAPHY

Y.A. Stepanyants[1,2]

[1]School of Agricultural, Computational and Environmental Sciences, University of Southern Queensland, QLD 4350, Australia, E-mail: Yury.Stepanyants@usq.edu.au;
[2]Department of Applied Mathematics, Nizhny Novgorod State Technical, University, Nizhny Novgorod, 603950, Russia.

Abstract

This paper presents specific features of solitary wave dynamics within the framework of the Ostrovsky equation with variable coefficients in relation to surface and internal waves in a rotating ocean with a variable bottom topography. For solitary waves moving toward the beach, the terminal decay caused by the rotation effect can be suppressed by the shoaling effect. Two basic examples of a bottom profile are analyzed in detail and supported by direct numerical modelling. One of them is a constant-slope bottom and the other is a specific bottom profile providing a constant amplitude solitary wave. Estimates with real oceanic parameters show that the predicted effects of stable soliton dynamics in a coastal zone can occur, in particular, for internal waves.

# 1. Introduction

As is well-known, the Korteweg–de Vries (KdV) equation, generalized to include the effect of the Earth rotation (the Ostrovsky equation) [1], does not have solutions in the form of stationary propagating solitary waves [2, 3]. Nevertheless, nonstationary solitary waves in the form of KdV solitons can propagate over long distances, experiencing a gradual decay due to the radiation of small-amplitude quasi-linear waves if the rotation effect is relatively small. In such cases a KdV soliton completely decays over a finite distance [4–6], whereas the radiation emitted by a soliton eventually transforms into an envelope soliton [7–12]. However, solitary waves can exist as stationary formations on the background of long periodic waves [13–16] or stably propagating along such waves periodically accelerating and decelerating, growing and decaying [16, 17].

In a non-homogeneous environment the dynamics of solitary waves can be more complex because of the interplay of several concurrent effects such as dissipation, inhomogeneity and possible vanishing of nonlinearity or dispersion. As a result, solitary waves propagating onshore can preserve the amplitude of surface or pycnocline displacement. It will be shown below, such preservation can happen within the adiabatic theory at the special bottom profile. To author's best knowledge, such a possibility has not previously been considered despite numerous studies of solitary wave dynamics in oceans with a variable environment (see [18–21] and references therein).

This paper is devoted to the specific analysis of adiabatic transformation of KdV solitons in a rotating fluid over a sloping bottom; it demonstrates that the terminal damping caused by wave radiation due to the rotation effect can be suppressed. In some cases, a soliton on a water surface can formally arrive at a beach before it completely vanishes; however, in fact, the adiabatic theory becomes invalid earlier. In the case of internal waves on the interface between two layers, a solitary wave can attain a point where the thickness of the layers becomes equal. At this point, the coefficient of nonlinearity becomes zero together with the soliton amplitude within the framework of adiabatic theory.

A certain interest can represent the case of a specific bottom profile on which the soliton amplitude remains constant upon propagation (whereas the fluid velocity varies to conserve the total wave energy flux). Such bottom profiles have been found for both surface and internal waves.

The paper is organized as follows. In Sec. 2 the general formula similar to that derived in Ref. [19] is obtained from the conservation of wave energy flux in the time-like Ostrovsky equation with variable parameters. In Sec. 3 the adiabatic solution for surface wave soliton propagation is considered. This section first studies a model of a constant bottom slope and demonstrates a competitive contribution of the effects of terminal damping and inhomogeneity for onshore propagation. A specific bottom profile is then found, providing a constant soliton amplitude in the course of its propagation toward the beach. In Sec. 4, a similar but more complicated analysis is applied to internal wave solitons in a two-layer fluid with the varying thickness of the lower layer. The analytical solutions are also obtained and analysed. In Sec. 5, the obtained adiabatic solutions are confirmed by direct numerical modeling within the framework of the variable coefficient Ostrovsky equation. In the Conclusion, the results obtained are briefly discussed and summarized.

## 2. Rotation modified KdV equation with variable coefficients

As mentioned above, the adiabatic solution for the internal solitons propagating in a rotating fluid with a spatially varying topography was first obtained in [19]. Here the similar results are briefly reproduced in a slightly different, but equivalent, form, and then they are used as the basis for the derivation of important outcomes for surface and internal waves.

Consider the Ostrovsky equation with the variable coefficients for the displacement of a water surface $\eta$:

$$\frac{\partial}{\partial x}\left(\frac{\partial \eta}{\partial t} + c\frac{\partial \eta}{\partial x} + \alpha\eta\frac{\partial \eta}{\partial x} + \beta\frac{\partial^3 \eta}{\partial x^3} + \frac{\eta}{2}\frac{dc}{dx}\right) = \gamma\eta, \quad (1)$$

For surface waves the coefficients of this equation are

$$c = \sqrt{gh}, \quad \alpha = \frac{3c}{2h}, \quad \beta = \frac{ch^2}{6}, \quad \gamma = \frac{f^2}{2c}, \quad (2)$$

where $h(x)$ in the fluid depth and $f$ is the Coriolis parameter [5]. For internal waves in a two-layer fluid, $\eta(x, t)$ stands for a displacement of a pycnocline, and the coefficients in the Boussinesq approximation are:

$$c = \sqrt{g \frac{\delta\rho}{\rho} \frac{h_1 h_2}{h_1 + h_2}}, \quad \alpha = \frac{3c}{2} \frac{h_1 - h_2}{h_1 h_2}, \quad \beta = \frac{c h_1 h_2}{6}, \quad \gamma = \frac{f^2}{2c}. \tag{3}$$

Here $h_{1,2}$ are thicknesses of the upper and lower layers, and it is assumed that $h_1$ = const., and $h_2 = h_2(x)$.

For the boundary-value (signaling) problem with $\eta(t, 0) = f(t)$, this equation can be presented in the alternative form dubbed here the time-like Ostrovsky equation:

$$\frac{\partial}{\partial t}\left(\frac{\partial \eta}{\partial x} + \frac{1}{c(x)}\frac{\partial \eta}{\partial t} - \frac{\alpha}{c^2(x)}\eta\frac{\partial \eta}{\partial t} - \frac{\beta}{c^4(x)}\frac{\partial^3 \eta}{\partial t^3} + \frac{\eta}{2c(x)}\frac{dc}{dx}\right) = -\gamma \eta. \tag{4}$$

In the absence of rotation in a fluid with constant parameters Eq. (4) has a family of solitary solutions which can be presented as:

$$\eta = A \operatorname{sech}^2\left(\frac{t - x/V}{T}\right), \tag{5}$$

where the soliton velocity $V$ and characteristic duration $T$ are linked with the amplitude $A$:

$$V = \frac{c}{1 - \alpha A/3c} \approx c + \frac{\alpha}{3}A, \quad T = \frac{\Delta}{c} = \frac{1}{c}\sqrt{\frac{12\beta}{\alpha A}}, \tag{6}$$

where $\Delta = \sqrt{12\beta/\alpha A}$ is the characteristic half-width of a soliton. We remind the reader that Eq. (4) and all subsequent formulae, as well as the Ostrovsky equation per se, are valid for weakly nonlinear perturbations; in particular, for a soliton it is required that $\alpha A/3c \ll 1$.

If the boundary condition for the time-like Ostrovsky equation (4) at $x = x_0$ is given in the form of a KdV soliton (5), then as known [4–6], it experiences adiabatic decay due to the influence of small radiation, provided that $\gamma \ll 1$. However, inhomogeneity can either enhance or diminish this effect, depending on the gradient of the background.

Multiplying Eq. (4) by $u$ and then integrating it over $t$, we obtain the energy balance equation in the form:

$$\frac{d}{dx}\int_{-\infty}^{+\infty} \eta^2 dt = -\gamma\left(\int_{-\infty}^{+\infty} \eta\, dt\right)^2 - \frac{1}{c}\frac{dc}{dx}\int_{-\infty}^{+\infty} \eta^2 dt, \tag{7}$$

where the quantity $E = \frac{1}{2}\int_{-\infty}^{+\infty} \eta^2 dt$ is usually treated as the "soliton energy" or wave action [19]. This quantity is proportional to the real wave energy, which follows from the primitive set of hydrodynamic equations in the long-wave approximation (see Appendix).

Substituting here the solution in the form of the KdV soliton (5), we derive the equation determining soliton amplitude variation with $x$:

$$\frac{dA}{dx} = -\frac{2\gamma}{c}\sqrt{\frac{12\beta}{\alpha}}A^{1/2} - \frac{A}{3}\frac{d}{dx}\ln\frac{\beta}{\alpha}. \tag{8}$$

Note that by integration of Eq. (4) over $t$, one can derive the equation of mass balance, which looks rather as the constraint in the case of Ostrovsky equation [1, 5, 26, 29]: $\int_{-\infty}^{+\infty} \eta dt = 0$. This equation does not allow any derivation of the equation for the variation of soliton amplitude in space, whereas Eq. (7) is not only a consequence of the heuristic approach, but also follows from the rigorous application of the asymptotic method as described, for example, in [18, 6].

Equation (8) can be readily solved (cf. [19]):

$$\frac{A(x)}{A_0} = \left[\frac{Y(x_0)}{Y(x)}\right]^{1/3}\left\{1 - \Delta(x_0)\int_{x_0}^{x}\frac{\gamma(\xi)}{c(\xi)}\left[\frac{Y(\xi)}{Y(\xi_0)}\right]^{2/3}d\xi\right\}^2, \tag{9}$$

where $Y(x) = \beta(x)/\alpha(x)$.

As mentioned above, the total mass of wave perturbation within Eq. (4) is zero. The total mass consists initially of the mass of a soliton $M_s(0)$ and the total mass of a "pedestal" which compensates the soliton mass. The latter can be readily calculated, using Eq. (5):

$$M_s(x) = \int_{-\infty}^{+\infty} \eta(x,t)dt = A(x)\int_{-\infty}^{+\infty} \text{sech}^2\left(\frac{t-x/V}{T}\right)dt = 2A(x)T(x) = \frac{4}{c(x)}\sqrt{\frac{3A(x)\beta(x)}{\alpha(x)}}. \tag{10}$$

Here, the relationship between the soliton amplitude and duration was used, as per Eq. (6). In the process of evolution soliton produces a shelf [18] and radiates a wave train [6]. The soliton amplitude varies with $x$, and accordingly its mass also varies, but the total mass of the soliton, shelf and radiated wave remains the same as at $x = 0$. This allows us to determine the total radiated mass from the soliton:

$$M_r(x) = M_s(x_0) - M_s(x) = \frac{4}{c(x_0)}\sqrt{\frac{3A_0\beta(x_0)}{\alpha(x_0)}} - \frac{4}{c(x)}\sqrt{\frac{3A(x)\beta(x)}{\alpha(x)}}. \tag{11}$$

Substituting here solution (9), we finally obtain (cf. [30]):

$$\frac{M_r(x)}{M_s(0)} = 1 - \sqrt{\frac{h(x)}{h(x_0)}}\left[1 - \sqrt{\frac{h(x_0)}{3A_0}}\frac{f^2}{g}\int_{x_0}^{x}\frac{h(\xi)}{h(\xi_0)}d\xi\right]. \tag{12}$$

Thus, within the adiabatic theory, the initial soliton mass ultimately transfers to the mass of the radiated wave.

The variation of "soliton energy" $E$ and energy flux $J = Ec$ with $x$ can be established when the variation of soliton amplitude with $x$ is determined. The quantity $E$ as defined here is treated conditionally as the wave energy in physical applications; it conserves in a homogeneous ocean, but it is not a Hamiltonian for the Ostrovsky equation [5, 26]. On the soliton solution (5) the energy is:

$$E(x) = \frac{1}{2}\int_{-\infty}^{+\infty}\eta^2 dt = \frac{4}{c(x)}\sqrt{\frac{\beta(x)}{3\alpha(x)}}A^{3/2}(x). \tag{13}$$

Then, using solution (9) for the dependence of soliton amplitude on $x$, we derive:

$$\frac{J(x)}{J(0)} \equiv \frac{E(x)c(x)}{E(0)c(x_0)} = \left\{1 - \Delta(x_0)\int_{x_0}^{x}\frac{\gamma(\xi)}{c(\xi)}\left[\frac{Y(\xi)}{Y(\xi_0)}\right]^{2/3}d\xi\right\}^3. \tag{14}$$

This formula represents the variation of energy flux in a rotating inhomogeneous medium; without rotation (when $\gamma \equiv 0$) this quantity conserves.

For a homogeneous medium, when $c$, $\alpha$, $\beta$, and $\gamma$ are constants, Eq. (9) reduces to the well-known formula for the terminal decay of KdV soliton [4–6]:

$$A = A_0\left(1 - \frac{\gamma\Delta}{c}x\right)^2. \tag{15}$$

According to this formula, a soliton completely decays (transforms to the radiated wave train) at a finite distance $X_e = c/\gamma\Delta$. As shown in the papers [7–12], after a long evolution the radiation of a KdV soliton can transfer into a stationary moving nonlinear wave train – an envelope soliton. This transformation is not considered here, because it is assumed that the inhomogeneity effect manifests itself at shorter distances.

The general solution (9) indicates the potential existence of singularities. Indeed, according to Eqs. (2) and (3), at the beach $Y \to \infty$, so that the factor before the curly parentheses in Eq. (9) diverges. Another singularity appears in the two-layer case; if initially $h_2 > h_1$ is at some point far from the beach, but then $h_2$ decreases approaching the beach, it will be a point where $h_2 = h_1$ and therefore $\alpha = 0$ as per Eq. (3). Soliton amplitude turns to zero in such a point. One more singularity can arise when the terminal damping occurs. This can happen when the integral in the curly parentheses grows and attains $1/\Delta(x_0)$; this depends on specific bottom profiles, which will be considered below.

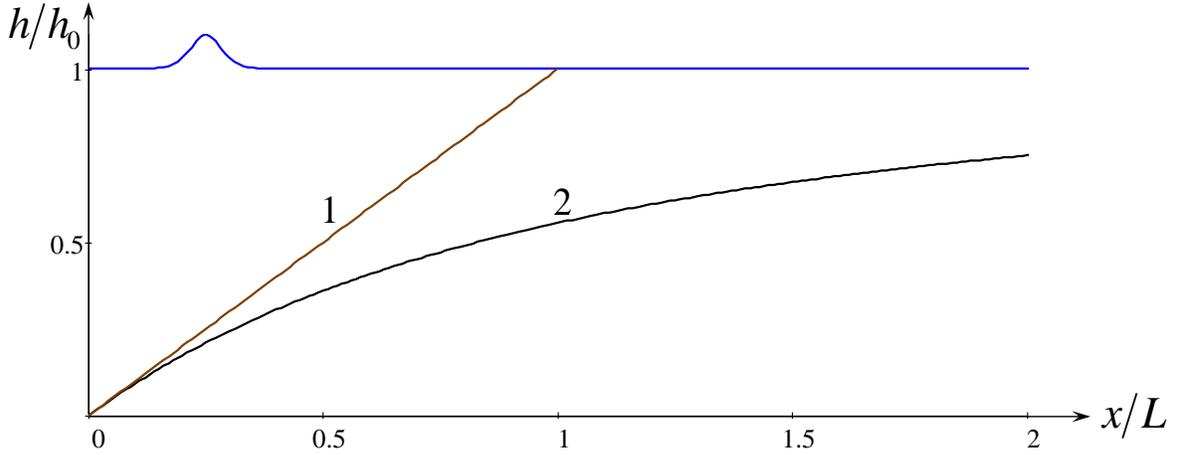

Fig. 1. Bottom profiles $h/h_0$ as functions of distance for the case of (i) constant bottom slope (line 1) and (ii) special bottom profile (will be discussed in subsection 3.2). A surface solitary wave is shown schematically, not in scale.

## 3. Influence of the Earth's rotation on the dynamics of surface solitons in the basins with a decreasing depth

Let us consider some interesting specific applications of the general formulae to the surface waves over a sloping bottom. We will consider two examples of bottom profile: (i) the constant sloping bottom, $h(x) = h_0(1 - x/L)$ (see line 1 in Fig. 1) and (ii) a special bottom profile, $h(x) = h_0(1 + x/2L)^{-2}$ (see line 2 in Fig. 1), which provides soliton propagation with a constant amplitude of surface displacement. The results obtained in this Section are rather model aiming to illustrate the solitary wave dynamics with the help of relatively simple expressions. More realistic physical examples will be considered in the next Section with application to internal waves, where the corresponding formulae are not so transparent.

### 3.1. Dynamics of a KdV soliton over a bottom with a constant slope

In this subsection it is assumed for simplicity that the bottom has a constant slope, $h(x) = h_0(1 - x/L)$, where $L$ is the characteristic distance of bottom variation (see line 1 in Fig. 1). In the considered case of the constant-slope bottom, the spatial variations of coefficients in Eq. (1) are (see Fig. 2):

$$c = \sqrt{gh_0}\left(1-\frac{x}{L}\right)^{1/2}, \quad \alpha = \frac{3}{2}\sqrt{\frac{g}{h_0}}\left(1-\frac{x}{L}\right)^{-1/2}, \quad \beta = \frac{1}{6}\sqrt{g}h_0^{5/2}\left(1-\frac{x}{L}\right)^{5/2}, \quad \gamma = \frac{f^2}{2\sqrt{gh_0}}\left(1-\frac{x}{L}\right)^{-1/2}. \quad (16)$$

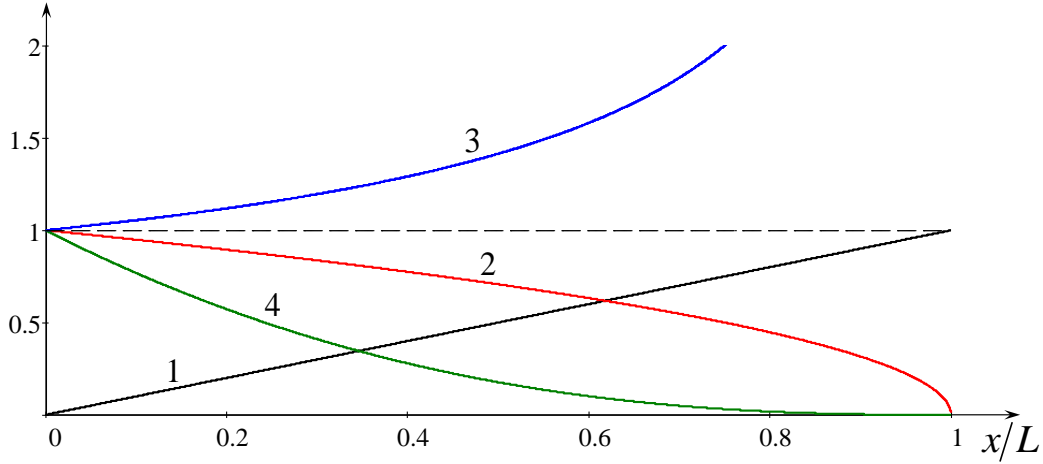

Fig. 2. Spatial variation of parameters in Eq. (2) for surface waves propagating in a water with linearly varying bottom profile. Line 1 shows $h(x)/h(0)$; line 2 – $c(x)/c(0)$; line 3 – $\alpha(x)/\alpha(0)$ and $\gamma(x)/\gamma(0)$; line 4 – $\beta(x)/\beta(0)$. The plot was generated for $h_0 = 500$ m and $L = 7.6 \cdot 10^7$ m.

Then we have in Eq. (9): $Y(x) = (1/9)h_0^3(1-x/L)^3$ and $\gamma(x)/c(x) = \left(f^2/2gh_0\right)(1-x/L)^{-1}$. Here the parameter $L$ represents the distance from the initial point $x_0 = 0$ to the beach. Substituting these expressions into Eq. (9), we obtain for the soliton amplitude:

$$\frac{A(x)}{A_0} = \left(1 - \frac{x}{L}\right)^{-1} \left[1 - \frac{f^2}{g}\sqrt{\frac{h_0}{3A_0}} x \left(1 - \frac{x}{2L}\right)\right]^2, \qquad (17)$$

where $A_0 = A(0)$. If there is no rotation ($f = 0$), then the normalized soliton amplitude $A(x)/A_0$ does not depend on the initial amplitude $A_0$ and gradually increases with distance in accordance with the hyperbolic law – see line 1 in Fig. 3.

It is easy to verify that in the limit of a flat bottom, $L \to \infty$, this formula reduces to Eq. (15) describing terminal decay of a KdV soliton (see line 2 in Fig. 3). Variations of soliton amplitude upon onshore propagation for different initial amplitudes are shown in Fig. 3.

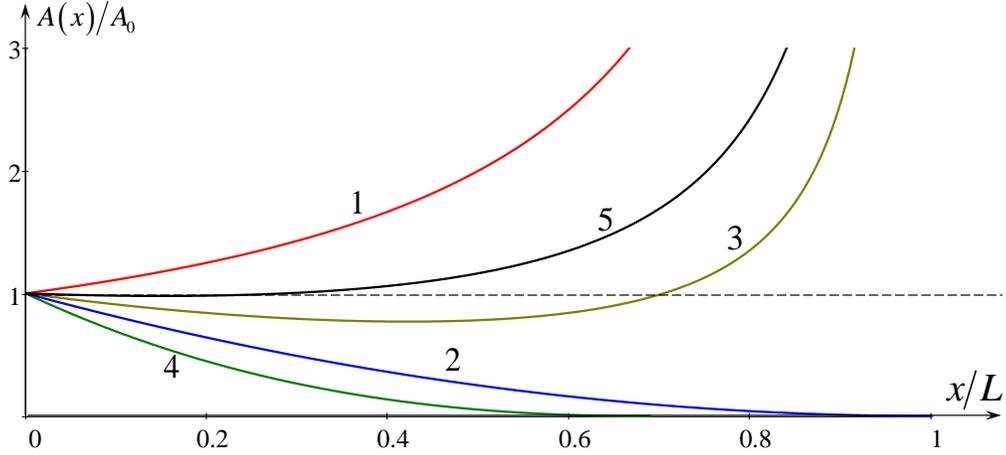

Fig. 3. Spatial variation of KdV soliton amplitude with distance as per Eq. (15) for different initial amplitudes. Line 1 shows the amplitude dependence when the rotation is absent ($f = 0$); line 2 pertains to the case when a soliton of initial amplitude $A_0 = 1$ m experiences a terminal decay in a basin with a flat bottom; line 3 – the same as in line 2, but in the basin with linearly increasing bottom; line 4 – the same as in line 3, but for a soliton of amplitude $A_0 = 0.2$ m; and line 5 – the same as in line 3, but for a soliton of amplitude $A_0 = 2.5$ m. The plot was generated for $h(0) = 500$ m, $L = 7.6 \cdot 10^4$ km, and $f = 10^{-4}$ s$^{-1}$.

The analysis of Eq. (17) demonstrates that in the case of onshore propagation a soliton can attain a beach only if its amplitude is big enough. In the case of a flat bottom this can occur if $A_0 \geq A_{cr} \equiv h_0 f^4 L^2 / 3g^2$. In particular, if we set $h_0 = 500$ m, $L = 7.6 \cdot 10^4$ km, and $f = 10^{-4}$ s$^{-1}$, then we obtain that solitary waves of initial amplitudes $A_0 \geq 1$ m attain a beach (all these parameters except $L$ are quite realistic); line 2 in Fig. 3 illustrates such a case, when the terminal decay formally occurs for a soliton of $A_0 = 1$ exactly at $x = L$. Solitons of smaller amplitudes do not reach a beach in the ocean with a flat bottom, as they completely vanish within the framework of asymptotic theory (actually, they transfer into envelope solitons after long-term evolution [7–12], as has been mentioned above). However, in the ocean with a linearly increasing bottom, solitons of even smaller initial amplitudes can attain a beach. Line 3 in Fig. 3 shows that a soliton of initial amplitude of 1 m does not experience terminal decay at the given parameters. The terminal decay can occur at such parameters if the initial soliton amplitude is small enough (see, for example, line 4 in Fig. 3 for $A_0 = 0.2$ m). As follows from the formula for the critical soliton amplitude, $A_{cr}$ quickly decreases, when $L$ decreases. In particular, if we set $L = 3.8 \cdot 10^4$ km, which is two times less than was chosen above, we conclude that solitary waves of initial amplitudes $A_0 \geq A_{cr} = 0.25$ m are capable attaining a beach in the ocean with a flat bottom.

At a certain condition a soliton can keep its initial amplitude nearly unchanged for a relatively long distance (see, for example, line 5 in Fig. 3 for $x < 0.4L$). This suggests the possibility that a bottom profile exists, which allows a soliton amplitude to be unchanged with distance, at least formally, until the asymptotic theory remains valid. In the next Section we will show that such a possibility can indeed occur, and here we present the formulae for the mass of radiated wave train, soliton energy, and energy flux as per Eqs. (12)–(14). The mass of radiated wave train monotonically increases in accordance with the formula (see line 1 in Fig. 4):

$$\frac{M_r(x)}{M_s(0)} = 1 - \sqrt{1 - \frac{x}{L}} \left[ 1 - \frac{f^2}{g} \sqrt{\frac{h_0}{3A_0}} x \left(1 - \frac{x}{2L}\right) \right]. \tag{18}$$

The energy flux in this case monotonically decreases due to the influence of rotation (see line 2 in Fig. 4), whereas without rotation it would be constant:

$$\frac{J(x)}{J(0)} = \left[ 1 - \frac{f^2}{g} \sqrt{\frac{h_0}{3A_0}} x \left(1 - \frac{x}{2L}\right) \right]^3. \tag{19}$$

And the wave energy, as defined in Eq. (13), varies with $x$ non-monotonically; it decreases first, but then abruptly increases when soliton approaches a beach (see line 3 in Fig. 4):

$$\frac{E(x)}{E(0)} = \frac{1}{\sqrt{1 - x/L}} \left[ 1 - \frac{f^2}{g} \sqrt{\frac{h_0}{3A_0}} x \left(1 - \frac{x}{2L}\right) \right]^3. \tag{20}$$

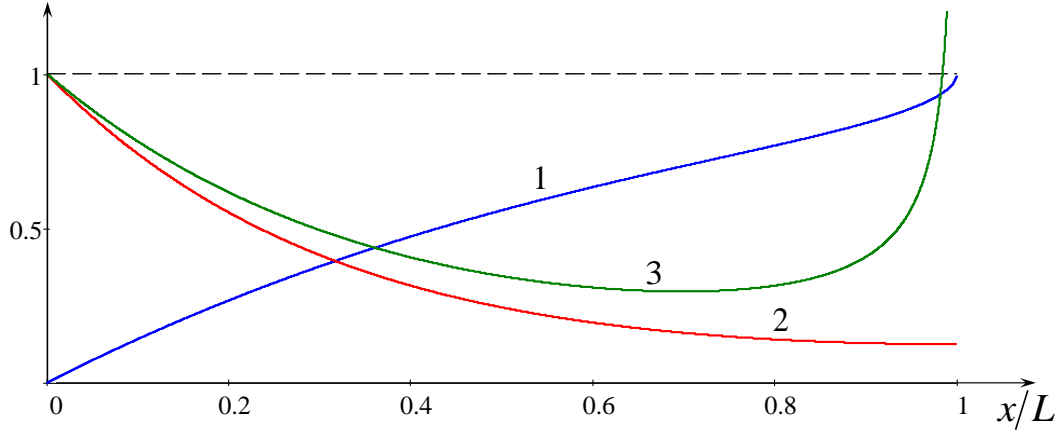

Fig. 4. Spatial variation of wave mass transferred from the KdV soliton of initial amplitude $A_0 = 1$ m to the trailing wave $M_r(x)/M_r(0)$ (line 1), the energy flux associated with the KdV soliton $J(x)/J(0)$ (line 2), and the soliton energy $E(x)/E(0)$ (line 3). All parameters are the same as in Fig. 3.

### 3.2. *Dynamics of a KdV soliton with a constant amplitude over a special bottom profile*

As follows from Eq. (9), the effects of shoaling and rotation can compensate for each other at certain conditions. In such special cases, a solitary wave can propagate onshore keeping the amplitude unchanged. However, the energy of the soliton gradually decreases due to the permanent radiation of small-amplitude waves caused by the rotation effect [4–6].

Let us calculate the bottom profile, which allows a KdV soliton to keep its amplitude unchanged. Equating the left-hand side of Eq. (9) to unity and differentiating the resultant equation with respect to $x$, we obtain (see line 2 in Fig. 1):

$$\frac{h(x)}{h_0} = \left(1 + \frac{\gamma(0)\Delta_0}{c(0)} x\right)^{-2} = \left(1 + \frac{x}{2L}\right)^{-2}, \qquad (21)$$

where $L = gh_0/f^2\Delta_0$. For relatively small distances, $x \ll 2L$, the bottom profile (21) reduces to the same linear profile as in subsection 3.1, whereas for $x \gg 2L$, we obtain from Eq. (21) $h(x) \approx h_0(2L/x)^2$. For $h_0 = 500$ m, $f = 10^{-4}$ s$^{-1}$, and initial soliton amplitude $A_0 = 1$ m ($\Delta_0 \approx 12.9$ km), we obtain $L = 3.8 \cdot 10^4$ km. This estimate agrees with what was obtained for the linear bottom profile and shown in Fig. 3 by line 5 – on the distances up to $x \approx L/3$ soliton amplitude remains almost unchanged.

The spatial dependences of coefficients (2) in the Ostrovsky equation (1) for this special case are shown in Fig. 5 (cf. Fig. 2). The spatial dependence of soliton amplitude in the course of soliton propagation onshore over the bottom profile (21) is described by the following equation:

$$\frac{A(x)}{A_0} = \frac{h_0}{h(x)}\left[1 - \frac{2Lf^2}{g}\sqrt{\frac{h_0}{3A_0}}\left(1 - \sqrt{\frac{h(x)}{h_0}}\right)\right]^2, \quad (22)$$

where $A_0$ is the soliton amplitude at the point where the thickness of the lower layer is $h_0 \equiv h(0)$.

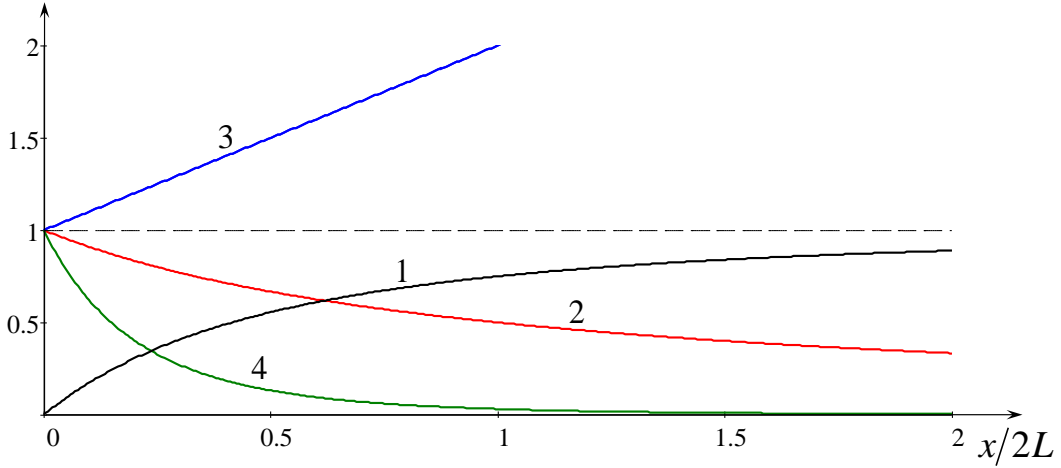

Fig. 5. Spatial variation of parameters as per Eq. (2) for surface waves propagating in a water with a specific bottom profile (21). Line 1 shows $1 - h(x)/h(0)$; line 2 – $c(x)/c(0)$; line 3 – $\alpha(x)/\alpha(0)$ and $\gamma(x)/\gamma(0)$; line 4 – $\beta(x)/\beta(0)$. The plot was generated for $h_0 = 500$ m and $L = 3.8 \cdot 10^7$ m.

Figure 6 demonstrates the variations of soliton amplitude upon onshore propagation over the bottom profile (21) for different initial amplitudes. If $A_c = 4L^2f^4h_0/3g^2$, then the soliton amplitude remains constant (see dashed line 3 in the figure). Line 1 shows the amplitude dependence $A \sim h^{-1}(x)$ when the rotation is absent ($f = 0$); line 2 pertains to the case when a soliton of initial amplitude $A_0 = A_c$ experiences a terminal decay due to the rotation in a basin with a flat bottom [5–6]; line 4 pertains to the case when a soliton of initial amplitude $A_0 = 2A_c$ travels onshore in a rotating ocean over the bottom profile (21); and line 5 pertains to the similar case as in line 4, but for a soliton of initial amplitude $A_0 = 0.5A_c$.

We can calculate again the variations with $x$ of radiated wave train mass, soliton energy, and energy flux as per Eqs. (12)–(14). These quantities are of a special interest for a soliton propagating with a constant amplitude. Then according to Eq. (12), the soliton permanently radiates a wave train, whose mass monotonically increases in accordance with the formula:

$$\frac{M_r(x)}{M_s(0)} = 1 - \left(1 + \frac{x}{2L}\right)^{-1}. \qquad (23)$$

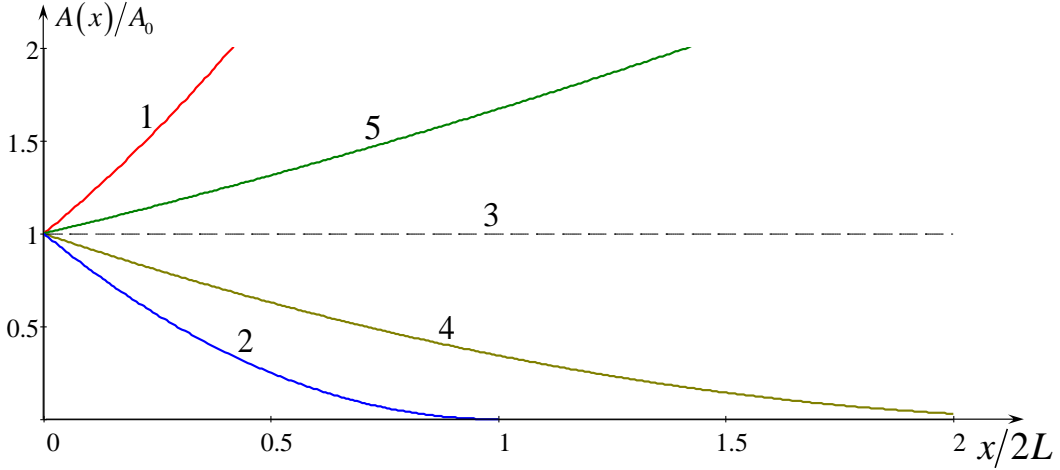

Fig. 6. Spatial variation of KdV soliton amplitude with distance. Line 1 shows the amplitude dependence $A \sim h^{-1}(x)$ when the rotation is absent ($f = 0$); line 2 pertains to the case when a soliton of initial amplitude $A_0 = A_c$ experiences a terminal decay due to rotation in a basin with a flat bottom; line 3 shows constant soliton amplitude when its initial amplitude $A_0 = A_c$; line 4 pertains to the case when a soliton of initial amplitude $A_0 = 2A_c$ travels in a rotating ocean over the bottom profile (20); and line 5 pertains to the similar case as in line 4, but for a soliton of initial amplitude $A_0 = 0.5A_c$. The plot was generated for $h(0) = 500$ m, $L = 3.8 \cdot 10^4$ km, and $f = 10^{-4}$ s$^{-1}$; for these parameters $A_c = 1$ m.

The energy flux monotonically decreases:

$$\frac{J(x)}{J(0)} = \left[\frac{h(x)}{h(x_0)}\right]^{3/2} = \left(1 + \frac{x}{2L}\right)^{-3}. \qquad (24)$$

The wave energy now decreases monotonically in contrast to the previous case described in subsection 3.1:

$$\frac{E(x)}{E(0)} = \frac{h(x)}{h(x_0)} = \left(1 + \frac{x}{2L}\right)^{-2}. \qquad (25)$$

Figure 7 illustrates the quantities $M_r(x)/M_r(0)$, $J(x)/J(0)$, and $E(x)/E(0)$ as functions of $x$ for the special case of soliton propagating with constant amplitude.

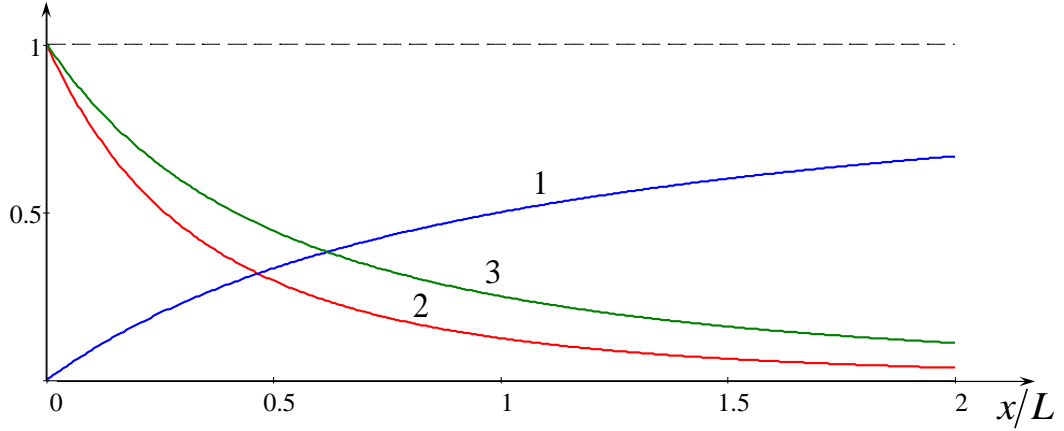

Fig. 7. Spatial variation of wave mass transferred from the KdV soliton of constant amplitude $A_c = 1$ m to the trailing wave $M_r(x)/M_r(0)$ (line 1), the energy flux associated with the KdV soliton $J(x)/J(0)$ (line 2), and the soliton energy $E(x)/E(0)$ (line 3). All parameters are the same as in Fig. 6.

While the amplitude of a solitary wave of amplitude $A_c$ remains unchanged, its width gradually decreases in the course of propagation toward the beach: $\Delta_c(x) = \Delta_c(0)(1 + x/2L)^{-3}$. As a result, the dispersion becomes not so small starting from a certain distance. The dispersion effect can be estimated as the ratio of the depth to soliton width: $h(x)/\Delta_c(x) = (3A_0/4h_0)^{1/2}(1 + x/2L)$. The nonlinearity (the ratio of soliton amplitude to the water depth) also increases because of the decreasing depth, $A_0/h(x) = (A_0/h_0)(1 + x/2L)^2$, therefore the small-amplitude approximation $A_0/h(x) \ll 1$ no longer holds.

Note that the applicability of the asymptotic theory, which requires a smallness of the rotation and shoaling effects in comparison with the nonlinear and dispersion effects, becomes better and better for the soliton of a special amplitude $A_c = $ const. moving onshore. Indeed, from an estimate of the rotation effect in comparison with the nonlinear effect it follows that $A_c \gg f h^{3/2}(x)/\sqrt{g}$, where $h(x)$ decreases in accordance with Eq. (21). Similarly from the estimate of the shoaling effect in comparison with the nonlinear effect we obtain $A_c \gg h^{5/3}(x)/L^{2/3}$.

A special example in this section, certainly, represents only a model due to an unrealistically big characteristic scale $L$ for the variation of bottom topography (whereas other parameters were taken to be quite realistic), nevertheless it still of interest from the general point of view as it helps to gain an insight into the physics of solitary wave dynamics under competing of shoaling and rotation effects. As will be shown below, a similar effect occurs in the more complicated case of internal waves in a two-layer rotating fluid, where a reasonable comparison with the realistic oceanic parameters is possible.

## 4. Internal solitons in a rotating two-layer ocean with a variable depth

In this Section we consider solitary waves propagating on the interface between two layers of stably stratified rotating fluid. The governing equation is the Ostrovsky Eq. (1) or its time-like counterpart Eq. (4) with the coefficients (3).

### 4.1. Dynamics of internal KdV soliton over a bottom with a constant slope

Let us assume again for simplicity that the bottom has a constant slope so that the thickness of the lower layer varies with $x$ as $h_2(x) = h_0(1 - x/L)$, whereas the thickness of the upper layer $h_1$ remains constant. Such a relatively simple model of linear bottom profile is nevertheless a typical approximation of a real topography in coastal zones (see, e.g., [19]). The spatial dependences of coefficients in Eq. (1) are shown in Fig. 8 for the particular set of hydrological parameters.

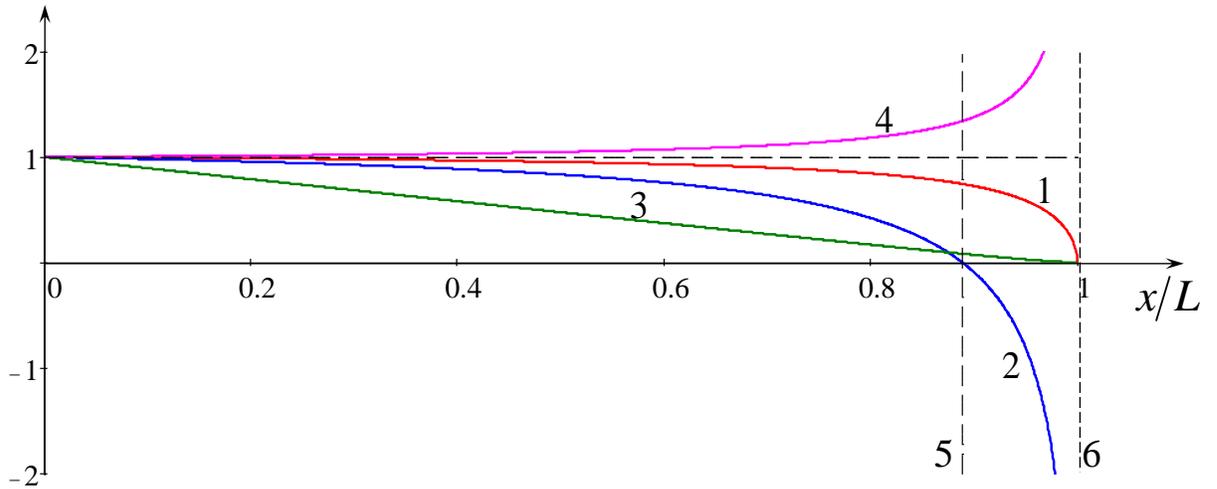

Fig. 8. Normalised dependences of coefficients of the Ostrovsky equation (1) for internal waves on distance for the linearly varying bottom. Line 1 shows $c(x)/c(0)$, line 2 – $\alpha(x)/\alpha(0)$; line 3 – $\beta(x)/\beta(0)$; line 4 – $\gamma(x)/\gamma(0)$. Dashed vertical line 5 shows the distance where $h_2 = h_1$, and dashed vertical line 6 shows the distance where $h_2 = 0$. The plot was generated for $h_1 = 50$ m, $h_0 = 450$ m, $f = 10^{-4}$ s$^{-1}$, $g' = 2.9 \cdot 10^{-2}$ m/s$^2$, and $L = 10^3$ km.

In this case, the solution to Eq. (9) can be formally presented in the analytic form:

$$\frac{A(x)}{A(0)} = \sqrt[3]{\frac{1 - x/L - h_1/h_0}{(1 - h_1/h_0)(1 - x/L)^2}} \left\{ 1 + 3\frac{\Delta_0}{\Lambda}\left(1 - \frac{h_1}{h_0}\right)^{2/3}\left(1 - \frac{x}{L} - \frac{h_1}{h_0}\right)^{1/3} \times \right.$$

$$\left. \left[ F\left(\frac{-1}{3}, \frac{1}{3}, \frac{4}{3}, 1 - \frac{x}{L}\frac{h_0}{h_1}\right) + F\left(\frac{-4}{3}, \frac{1}{3}, \frac{4}{3}, 1 - \frac{x}{L}\frac{h_0}{h_1}\right) \right]^2 \right\}, \qquad (26)$$

where $\Lambda = 2g'h_1/f^2L$, and $F(a, b, c, d)$ is the hypergeometric function.

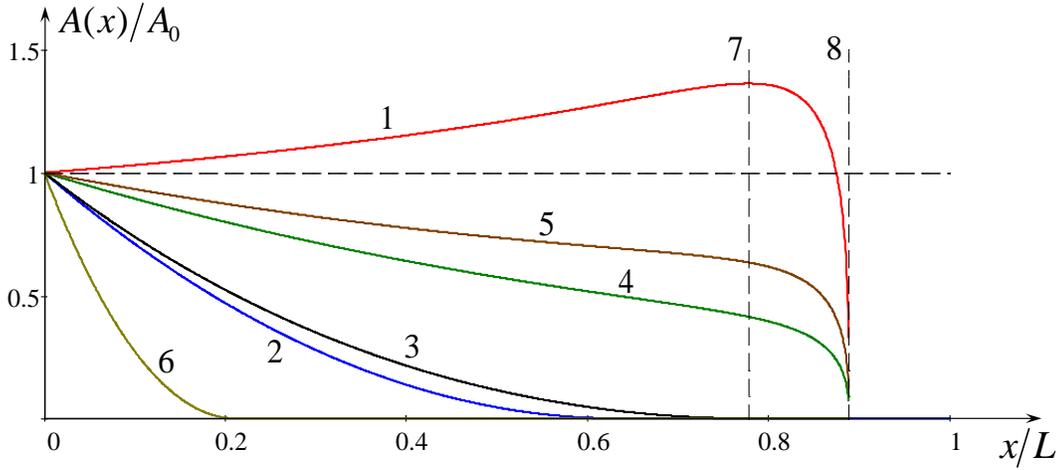

Fig. 9. Dependence of soliton amplitude on distance. Line 1 pertains to the case of non-rotating fluid; line 2 – to the case of rotating fluid with a flat bottom and $A_0 = 10$ m; line 3 – to the rotating fluid with a linearly varying bottom and $A_0 = 10$ m; line 4 – the same as line 3, but $A_0 = 50$ m; line 5 – the same as line 3, but $A_0 = 100$ m; line 6 – the same as line 3, but $A_0 = 1$ m. Dashed line 7 shows the distance where $h_2 = 2h_1$, and dashed line 8 shows the distance where $h_2 = h_1$. The plot was generated for $h_1 = 50$ m, $h_0 = 450$ m, $f = 10^{-4}$ s$^{-1}$, $g' = 2.9 \cdot 10^{-2}$ m/s$^2$, and $L = 10^3$ km.

In the non-rotating ocean ($f = 0$) this formula reduces to the earlier derived dependence of soliton amplitude on distance in the linearly varying bottom topography [21, 22] – see line 1 in Fig. 9. In such case, the amplitude of a solitary wave can be enhanced first if the thickness of the lower layer is greater than $2h_1$. Then it rapidly decreases and completely vanishes within the framework of adiabatic theory when $h_2$ decreases and becomes equal to $h_1$ (see the portion of line 1 between the vertical dashed lines 7 and 8). The maximum possible amplitude gain is attained at $h_2 = 2h_1$ and equals

$$\frac{A_{max}}{A(0)} = \left[4\frac{h_1}{h_0}\left(1-\frac{h_1}{h_0}\right)\right]^{-1/3}. \tag{27}$$

When a soliton approaches the position of the critical depth $h_2 = h_1$, it experiences a significant transformation. At this depth the nonlinear coefficient $\alpha$ in Eq. (1) vanishes (see Eq. (3)), and then becomes positive if $h_2 < h_1$. Soliton transformation at the critical point in a non-rotating fluid has been studied in many papers (see, e.g., [23, 24] and references therein). The change of sign of the nonlinear coefficient $\alpha$ implies a change of polarity for a solitary wave solution in the KdV equation. In the case $h_2 > h_1$, only solitons of negative polarity in the form of pycnocline depressions can exist, whereas in the case $h_2 < h_1$ only solitons of positive polarity in the form of pycnocline humps can exist. Therefore, when a soliton of negative polarity approaches a critical point where $\alpha = 0$, its further fate depends on the width of the transient zone. As was shown in Ref. [23], there are no solitary waves after wave passage through the critical point if the width of the transient zone is relatively short. However, if the width of the transient zone is very wide, much greater than the width of the approaching soliton, then one or more secondary solitons of positive polarity can appear after wave passage through the critical point.

In the case of a flat bottom, formula (26) provides the well-known result of terminal soliton decay [4–6] (see Eq. (15)). In the ocean with a sloping bottom, the extinction distance $X_e$ becomes greater for the soliton of the same initial amplitude as in the case of a flat bottom (cf. lines 3 and 2). But if the initial soliton amplitude decreases, and its widths increases, then the extinction distance decreases (cf. lines 6 and 3). In the opposite case, when the initial soliton amplitude increases (and its width decreases), the extinction distance increases and starting from a certain critical amplitude it completely disappears, because a soliton approaches a position where it is destroyed due to the vanishing of the nonlinear coefficient $\alpha$ at $h_2 = h_1$ (see lines 4 and 5). For the set of parameters used to generate the plots shown in Fig. 9 this occurs when the soliton amplitude attains 50 m which corresponds to a strongly nonlinear solitary wave with $A_0 =$ $h_0$. The Ostrovsky model is formally inapplicable to such strong waves, but fortunately, it still provides quite reasonable estimates even in such cases (see, for example, [25–27]).

A similar analysis can be carried out for the case when at the initial point $x = x_0$ the thickness of the lower layer is $h_2 < h_1$, and then decreases with $x$ when a soliton moves onshore (note that in this case soliton amplitude is positive, i.e. it looks like a hump on a pycnocline). Formula (26) is still applicable and provides the results for the soliton amplitude as the function of a distance shown in Fig. 10. This figure again clearly illustrates the competing effects of rotation and shoaling. In a non-rotating fluid soliton amplitude increases due to shoaling in accordance with the formula (see line 1 in the figure):

$$\frac{A(x)}{A(0)} = \left[\frac{h_1/h_0 - 1 + x/L}{(h_1/h_0 - 1)(1 - x/L)^2}\right]^{1/3}. \tag{28}$$

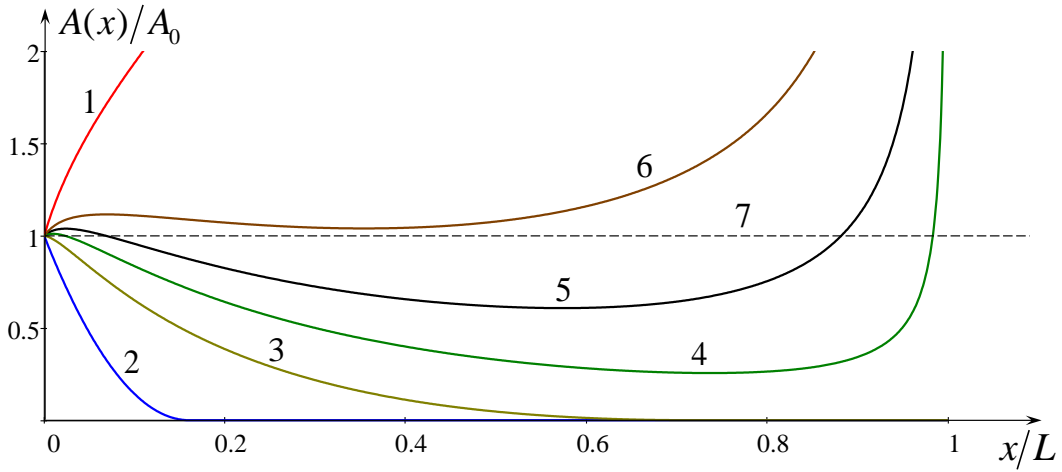

Fig. 10. Dependence of soliton amplitude on distance. Line 1 pertains to the case of non-rotating fluid as per Eq. (28); other lines pertain to a rotating fluid: line 2 – corresponds to the case a flat bottom and $A_0 = 10$ m; other lines 3–6 correspond to the linearly decreasing bottom and $A_0 = 5$ m (line 3); $A_0 = 7.5$ m (line 4); $A_0 = 10$ m (line 5); $A_0 = 15$ m (line 6). Dashed line 7 shows the reference case when soliton moves in the non-rotating fluid with a flat bottom. The plot was generated for $h_1 = 50$ m, $h_0 = 49$ m, $f = 10^{-4}$ s$^{-1}$, $g' = 2.9 \cdot 10^{-2}$ m/s$^2$, and $L = 10^3$ km.

In the rotating fluid with a flat bottom a soliton of initial amplitude $A_0 = 10$ m terminally decays at $X_{term} = 0.16L$ (see line 2), whereas a soliton of the same initial amplitude does not disappear in the fluid with a linearly increasing bottom (see line 5). Its amplitude slightly increases in the beginning, then gradually decreases, and, at last, abruptly increases when the thickness of the lower layer becomes very small. Similar behavior occurs with a soliton of $A_0 = 7.5$ m (see line 4), whereas a soliton of initial amplitude $A_0 = 5$ m experiences a terminal decay, but over a longer distance than in the flat bottom case (cf. lines 2 and 3). Solitons with the initial amplitudes $A_0 > 15$ m always have amplitudes greater than the initial one (see line 6).

*4.2. Dynamics of a KdV soliton with a constant amplitude over a special bottom profile*

The similar effect of a competition between the rotation and shoaling can be considered for internal waves in a two-layer fluid. In this case, using Eq. (9) and coefficients of the Ostrovsky equation (3), we obtain the equation for the depth profile:

$$\frac{dh_2}{dx} = \frac{2c(0)f^2}{g'}\sqrt{\frac{3(h_2-h_1)}{A_0}} h_2 \frac{h_1+h_2}{2h_2-h_1}, \qquad (29)$$

where $g' = g\delta\rho/\rho$ (see Eq. (3)).

The solution to this equation depends on the initial thickness of the lower layer $h_0$ when a soliton commences its motion toward the beach. If $h_2 > 2h_1$, then the solution in the implicit form is:

$$\frac{x-x_0}{L} = 3\left(\tan^{-1}\sqrt{\frac{h_0-h_1}{2h_1}} - \tan^{-1}\sqrt{\frac{h_2(x)-h_1}{2h_1}}\right) - 2\sqrt{2}\left(\tan^{-1}\sqrt{\frac{h_0-h_1}{h_1}} - \tan^{-1}\sqrt{\frac{h_2(x)-h_1}{h_1}}\right), \qquad (30)$$

where $L = (g'/f^2)(|A_0|/6h_1)^{1/2}$ is the characteristic scale of depth variation. This dependence is shown in Fig. 11 by line 1.

If the initial thickness of the lower layer is in the range $h_1 < h_2 < 2h_1$, then the solution in the implicit form is:

$$\frac{x-x_0}{L} = -3\left(\tan^{-1}\sqrt{\frac{h_0-h_1}{2h_1}} - \tan^{-1}\sqrt{\frac{h_2(x)-h_1}{2h_1}}\right) + 2\sqrt{2}\left(\tan^{-1}\sqrt{\frac{h_0-h_1}{h_1}} - \tan^{-1}\sqrt{\frac{h_2(x)-h_1}{h_1}}\right). \qquad (31)$$

This dependence is shown in Fig. 11 by line 2 with the choice of the arbitrary parameter, which provides matching of lines 1 and 2 at the level $h_2 = 2h_1$ (see dashed line 7).

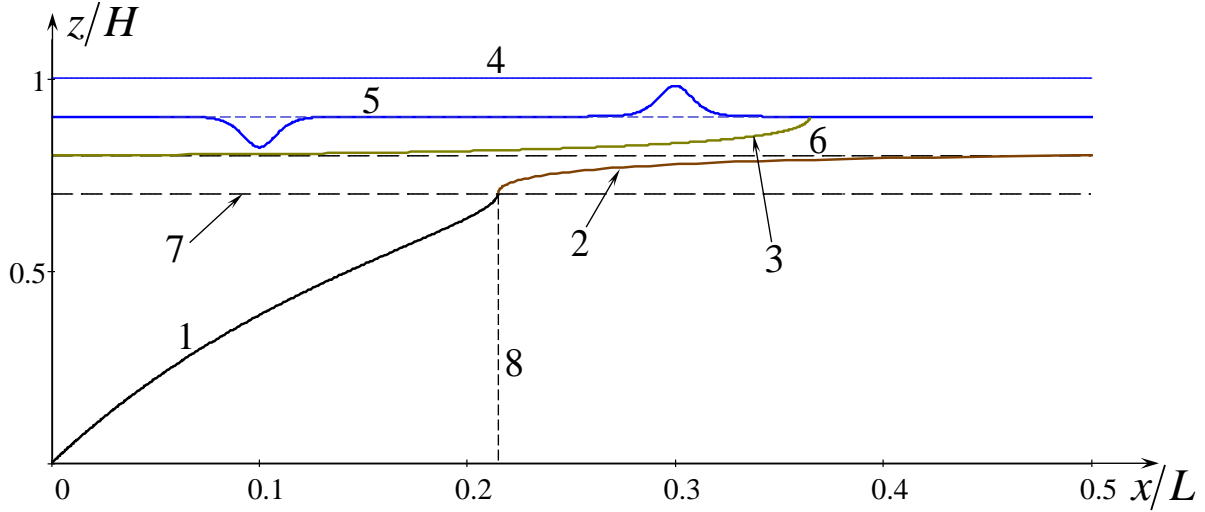

Fig. 11. Normalised depth $z/H$ where $H = h_1 + h_2(0)$ as a function of $x$ according to Eqs. (30) and (31). Line 1 pertains to the case when $h_2 > 2h_1$, line 2 – to the case when $h_1 < h_2 < 2h_1$, and line 3 – to the case when $0 < h_2 < h_1$. Line 4 shows a free surface, and line 5 – a pycnocline with the solitons of negative polarity, if $h_2 > h_1$, and positive polarity if $h_2 < h_1$. Line 6 shows the critical depth where $h_2 = h_1$, and the vertical dashed line 8 shows the distance where the bottom attains the level where $h_2 = 2h_1$ (see line 7).

As is well-known, solitary waves on the interface have negative polarities (depression waves as shown schematically in Fig. 11) if the lower layer is thicker than the upper layer; otherwise solitary waves have positive polarities. For such a hydrology configuration when $h_2 < h_1$ the bottom profile providing soliton propagation in a rotating fluid with the constant amplitude is also possible. It follows from the solution of Eq. (29) where $h_2$ and $h_1$ should be inter-replaced; the corresponding solution reduces again to Eq. (31) and is shown in Fig. 11 by line 3.

The dependences of coefficients in Eq. (1) on the distance for the specific bottom profiles are shown in Fig. 12. Vertical dashed lines show the distances where $h_2 = 2h_1$ (line 5), $h_2 = h_1$ (line 6), and $h_2 = 0$ (line 7).

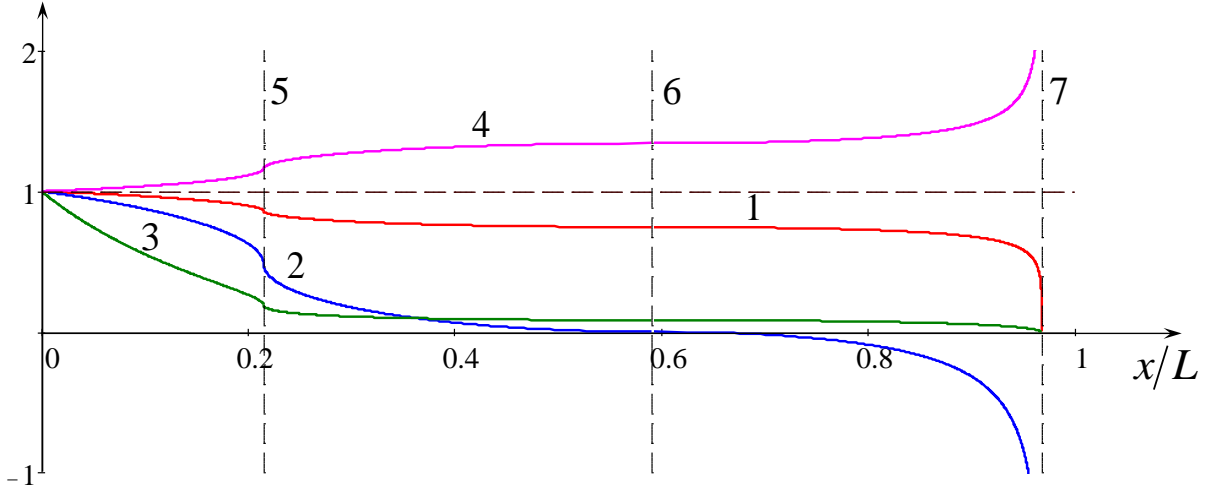

Fig. 12. Normalised dependences of coefficients of Ostrovsky equation (1) for internal waves on distance in a water with a specific bottom profile as per Eqs. (30) and (31). Line 1 shows $c(x)/c(0)$, line 2 – $\alpha(x)/\alpha(0)$; line 3 – $\beta(x)/\beta(0)$; line 4 – $\gamma(x)/\gamma(0)$. Dashed vertical line 5 shows the distance where $h_2 = 2h_1$, dashed vertical line 6 shows the distance where $h_2 = h_1$, and dashed vertical line 7 shows the distance where $h_2 = 0$. The plot was generated for $h_1 = 50$ m, $h_0 = 450$ m, $f = 10^{-4}$ s$^{-1}$, $g' = 2.9 \cdot 10^{-2}$ m/s$^2$, and $L = 10^3$ km.

In the particular case, when the amplitude of internal soliton remains constant in the course of propagation over the bottom of a special profile, its width varies as:

$$\Delta(x) = \sqrt{\frac{12\beta(x)}{\alpha(x)A_0}} = \frac{2h_1 h_2(x)}{\sqrt{3A_0 \left[h_1 - h_2(x)\right]}} \qquad (32)$$

This dependence in the normalized form $\dfrac{\Delta(x)}{\Delta(0)} = \dfrac{h_2(x)}{h_0}\sqrt{\left|\dfrac{h_1 - h_0}{h_1 - h_2(x)}\right|}$ is shown in Fig. 13 for three intervals of x-axis: (1) where $h_2 > 2h_1$ (line 1), (2) where $h_1 < h_2 < 2h_1$ (line 2), (3) where $0 < h_2 < h_1$ (line 3). Soliton width decreases at the first interval until it reaches the minimum value $\Delta(x) = 4\sqrt{2}\Delta(0)/9$.

Note that a similar effect of reflectionless propagation of linear internal waves in two-layer fluid was considered for specific bottom configurations (see [20] and other references therein). This allows internal waves travelling over large distances to transfer significant portions of momentum and energy toward the beach.

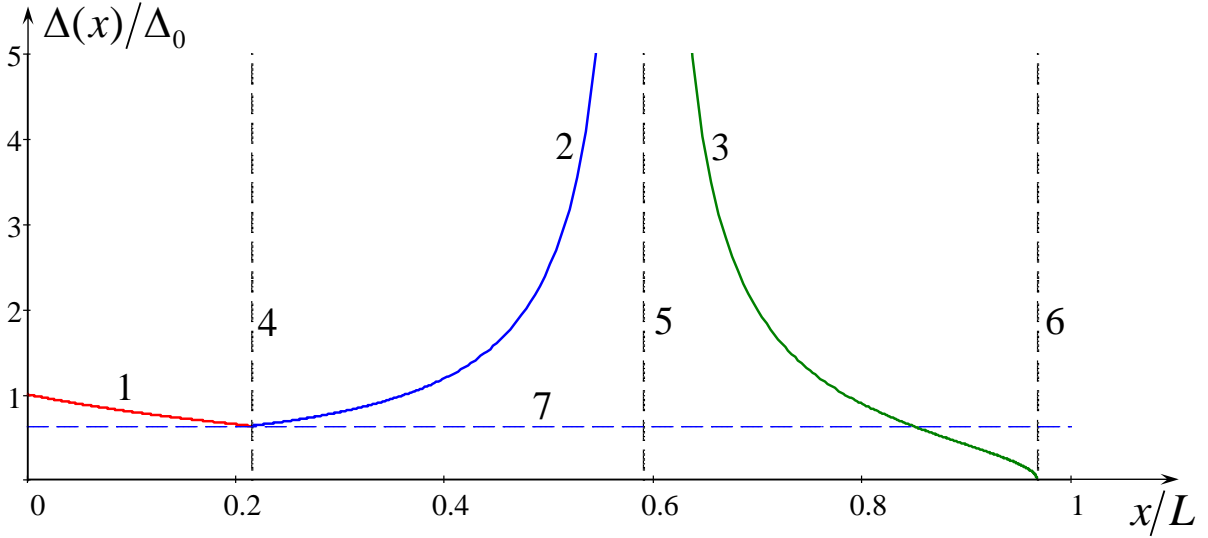

Fig. 13 (color online). Normalised dependences of soliton width for internal waves on distance in a water with a specific bottom profile as per Eqs. (30) and (31). Dashed vertical line 4 shows the distance where $h_2 = 2h_1$, dashed vertical line 5 shows the distance where $h_2 = h_1$, and dashed vertical line 6 shows the distance where $h_2 = 0$, dashed horizontal line 7 shows the minimal value of soliton width. The plot was generated for $h_1 = 50$ m, $h_0 = 450$ m, $f = 10^{-4}$ s$^{-1}$, $g' = 2.9 \cdot 10^{-2}$ m/s$^2$, and $L = 10^3$ km.

## 5. Numerical solution for soliton dynamics in inhomogeneous rotating fluid

From the numerical point of view, it is convenient to present Eq. (4) in the dimensionless form, following the well-known procedure (see, for example, [19]:

$$\frac{\partial}{\partial \tau}\left(\frac{\partial \upsilon}{\partial s} - \frac{\alpha(s)c^{3/2}(s)}{\beta(s)}\upsilon\frac{\partial \upsilon}{\partial \tau} - \frac{\partial^3 \upsilon}{\partial \tau^3}\right) = -\frac{\gamma(s)c^4(s)}{\beta(s)}\upsilon, \tag{33}$$

where $s = \int_0^x \frac{\beta(x')dx'}{c^4(x')}$, $\tau = t - \int_0^x \frac{dx'}{c(x')}$, $\zeta = \eta\sqrt{c(x)}$.

Equation (33) was numerically solved with the periodic boundary conditions on $\tau$ using a solitary wave (5) as the boundary condition at $s = 0$ (the numerical algorithm was described in Ref. [28]). Results of computations for surface waves are shown in Fig. 14; similar results were obtained for internal waves in a two-layer fluid.

First of all, it was validated that the numerical code provides the correct data for a KdV soliton moving toward a shore over the linearly decreasing bottom without the rotation effect. As one can see, the numerical data around line 1 (triangles) are in a good agreement with the theoretical prediction which follows from Eq. (17) with $f = 0$ (see line 1 in Fig. 3). The numerical data around line 2 (rhombuses) also agree well with the corresponding theoretical dependence for the case of rotating fluid with the linearly decreasing depth as per Eq. (17) and $A_0 = 2$ m. The numerical data shown by dots pertain to the case when soliton of a constant amplitude propagates over the bottom of a special profile as per Eq. (21) and $A_0 = 4$ m. In the latter case soliton amplitude remains constant, but its width decreases in accordance with the formula derived in Section 3.2, $\Delta_c(x) = \Delta_c(0)(1 + x/2L)^{-3}$.

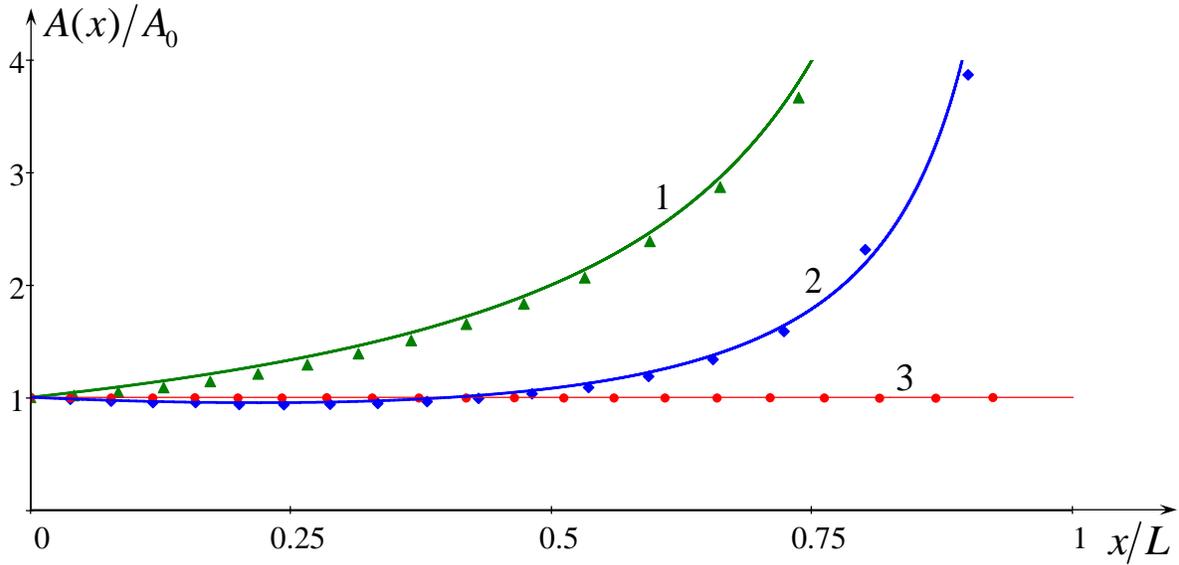

Fig. 14. Normalised soliton amplitude as a function of distance. Line 1 – theoretical dependence in the case of non-rotating fluid with the linearly decreasing depth; line 2 – theoretical dependence for the rotating fluid with the linearly decreasing depth and $A_0 = 2$ m, and line 3 is the constant amplitude predication for the case of special bottom profile as per Eq. (21) and $A_0 = 4$ m. Symbols represent numerical data for the corresponding theoretical dependences. The plot was generated for $h(0) = 500$ m, $L = 7.6 \cdot 10^4$ km, and $f = 10^{-4}$ s$^{-1}$.

In Fig. 15 one can see a comparison of initial surface soliton (line 1) at the position where the total depth was $H = 500$ m with the soliton at the positions where the total depth has decreased up to $H = 329$ m (line 2), and then further decreased up to $H = 225$ m (line 3). If the bottom was plane, then the soliton would disappear travelling over the same distance. However, travelling over the bottom of a special profile as per Eq. (21), the soliton keeps its amplitude, but noticeably shrinks, as predicted. In the tail part of a soliton shown by line 2 one can see formation of a secondary soliton which becomes well pronounced in line 3. This is a typical process accompanying terminal decay of KdV solitons [29, 6, 8] which eventually ends up with the formation of an envelope soliton in the fluid with a flat bottom [7–12]. In the fluid with the decreasing bottom, formation of envelope solitons at certain conditions can be suppressed by the shoaling effect. However, the study of long-term evolution of solitons and the possibility of envelope soliton formation in a fluid with an uneven bottom is beyond the scope of this paper.

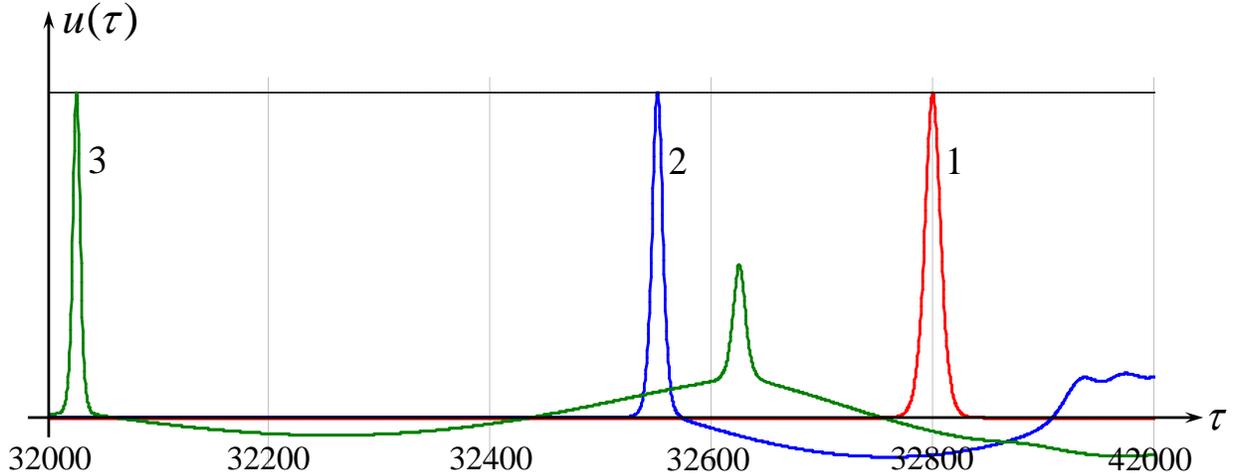

Fig. 15. Soliton profiles at different positions. Line 1 shows the initial KdV soliton at the depth $H = 500$ m; line 2 shows the soliton travelling over some distance where the total depth is $H = 329$ m, and line 3 shows the soliton travelling further for the same distance where the total depth decreases up to $H = 225$ m.

The asymptotic theory used in this paper is applicable, when the perturbation terms in the generalized KdV equation (1) proportional to $\gamma$ and $dc/dx$ are relatively small in comparison with the nonlinear and dispersive terms (which, in turn, are assumed to be small in comparison with the first two terms). Therefore, the developed here theory is applicable to solitary waves of relatively small or moderate amplitudes. In the meantime, it is of interest to investigate soliton behavior in the cases, when the perturbative terms are not so small, but comparable with the nonlinear and dispersive terms. The non-adiabatic evolution of a solitary wave in a coastal zone with a relatively strong bottom inhomogeneity has been studied numerically. Figure 16 demonstrates the dependences of soliton amplitude with the initial value $A_0 = 2$ m for several bottom slopes in the model with the linearly decreasing depth starting from $h(0) = 500$ m (see line 1 in Fig. 1); the Coriolis parameter was fixed, $f = 10^{-4}$ s$^{-1}$. Lines 1 and 2 in this figure are the same as in Fig. 14; they are shown here again as the reference plots for the sake of comparison with the cases when the asymptotic theory is applicable. Line 1 pertains to the case, when there is no fluid rotation; triangles are the numerical data. Line 2 is the theoretical dependence provided by the asymptotic theory with the numerical data (rhombuses) for the rotating fluid with a very small bottom slope ($L = 7.6 \cdot 10^4$ km). Line 3 represents the interpolation of numerical data (boxes) in the case of rotating fluid with a moderate bottom slope ($L = 7.6 \cdot 10^3$ km), but still beyond the asymptotic theory, and line 4 is the same as in line 3, but with a big bottom slope ($L = 7.6 \cdot 10^2$ km).

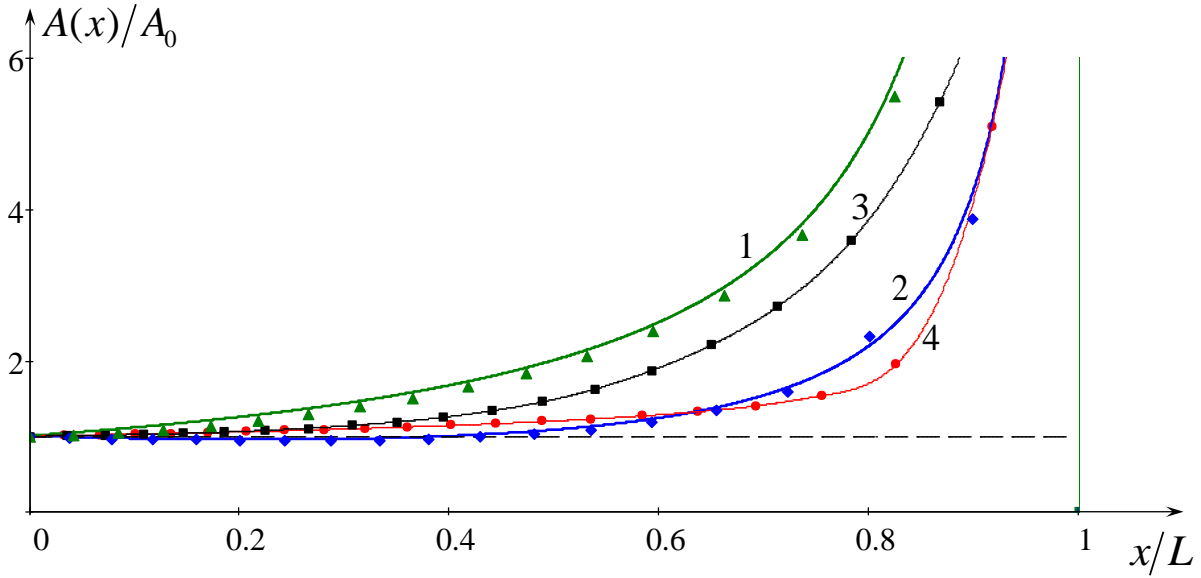

Fig. 16. Normalised soliton amplitude as a function of distance. Line 1 – theoretical dependence in the case of non-rotating fluid with the linearly decreasing depth; line 2 – theoretical dependence for the rotating fluid with the linearly decreasing depth and $L = 76{,}000$ km, line 3 – the best feet line for the numerical data in the case of linearly decreasing depth with $L = 7{,}600$ km in the rotating fluid, and line 4 is the same as line 3, but with $L = 760$ km. Symbols represent numerical data. The plot was generated for $h(0) = 500$ m, $f = 10^{-4}$ s$^{-1}$, and the initial soliton amplitude is $A_0 = 2$ m.

As follows from this figure, soliton amplitude monotonically increases upon approaching the coast. In the case of a moderate bottom slope, the amplitude dependence on distance (line 3) is qualitatively similar to the prediction of asymptotic theory for the non-rotating (line 1) and rotating (line 2) fluids. However, in the case of a big bottom slope, the character of amplitude dependence on distance is different (see line 4); under the synergetic action of rotation and bottom topography, the amplitude changes very slowly first, and then it abruptly increases

When the bottom slope is moderate ($L = 7{,}600$ km), the solitary wave profile remains qualitatively the same as in the case of a very small slope (see Fig. 17a). However, in the case of a big slope ($L = 760$ km), the incoming solitary wave breaks into a number of secondary solitary waves as shown in Fig. 17b). Line 4 in Fig. 16 shows the amplitude of only the first solitary wave growing with distance when it approaching the coast.

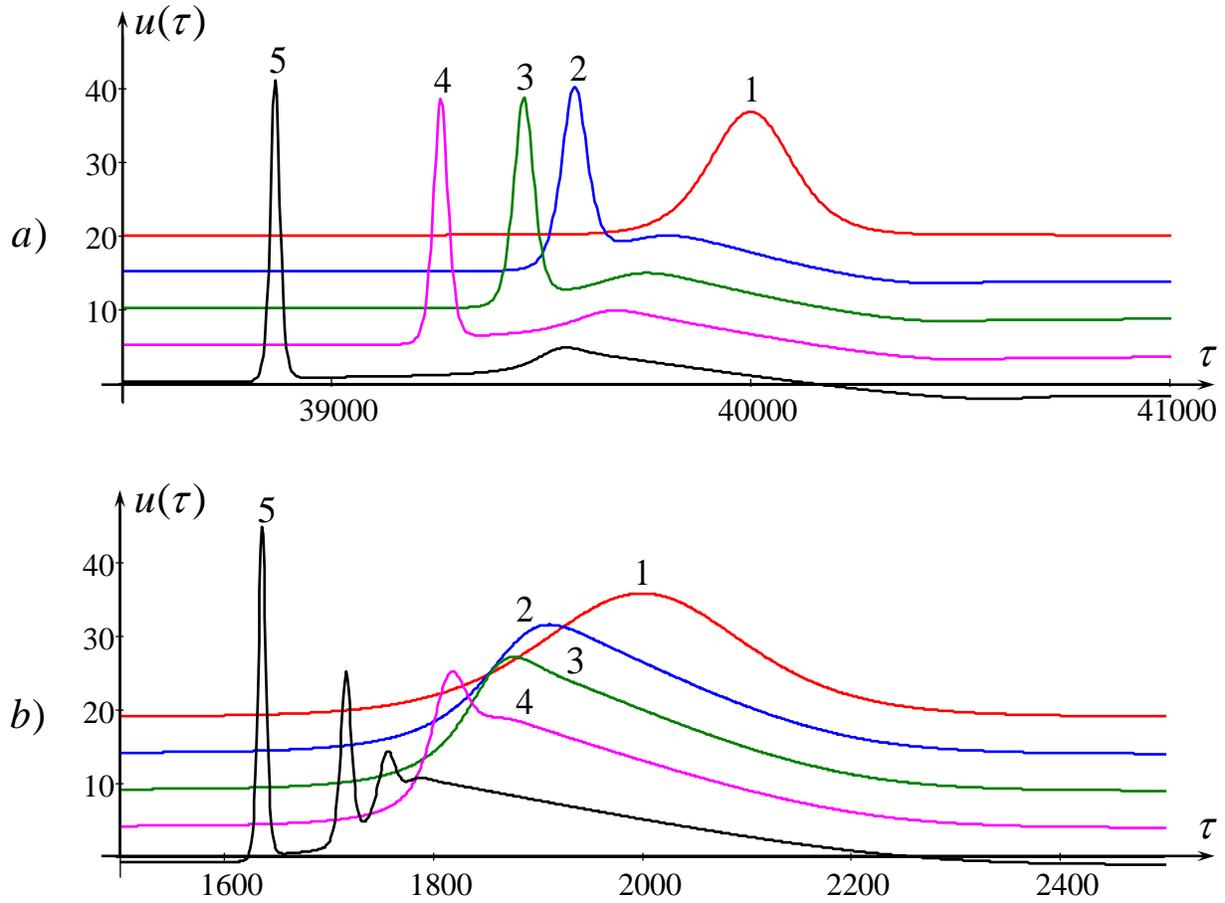

Fig. 17. Soliton profiles at different positions. Lines 1 in both frames show the initial KdV soliton at the depth $H = 500$ m. In frame a) solitary wave profiles are shown at the depths $H = 203.5$ m (line 2), 174.8 m (line 3), 143.5 m (line 4), 108.4 m (line 5). In frame b) solitary wave profiles are shown at the depths $H = 153.5$ m (line 2), 122.5 m (line 3), 86.9 m (line 4), 41.1 m (line 5).

## 6. Discussion and conclusion

The main aim of this paper was to demonstrate the specific interplay of two effects, fluid rotation and shoaling, on the dynamics of solitary waves in coastal zones. It has been shown that even within the framework of a relatively simple model of the Ostrovsky equation with the linearly decreasing bottom, some interesting and non-trivial effects can occur. In particular, the shoaling effect can suppress the terminal decay of solitary waves caused by the effect of the Earth's rotation.

It was suggested that with some special arrangements, solitary waves can propagate toward the beach with a constant amplitude of surface or pycnocline displacement. This idea has been confirmed and the corresponding conditions relating to the bottom profile and initial soliton amplitude have been found for both the surface and the internal waves. Note that, as follows from the linearised Euler equation, the particle velocity in a long surface wave is $u \sim g\eta/c(x)$. Therefore in the case of a soliton moving with a constant amplitude $A_0$ over the bottom of the special profile as per Eq. (21), the particle speed increases as $u \sim A_0\sqrt{g/h(x)} \sim A_0(1 + x/2L)\sqrt{g/h_0}$. All theoretical derivations have been validated by direct numerical modelling within the framework of the Ostrovsky equation with the variable coefficients. The estimates for the applicability of asymptotic theory used in this paper have been presented.

The numerical modelling within the framework of original Ostrovsky equation has confirmed the theoretical predictions for the relatively small bottom slope, when the asymptotic theory is applicable. It was found a very good agreement between the theoretical outcomes and numerical data in such cases, both for the constant slope bottom profile and for the special bottom profile, which provides soliton propagation with a constant amplitude. However, the range of applicability of asymptotic theory for oceanic waves can be very limited (especially for surface waves), because it requires relatively small bottom slopes and, as the result, very long distances for the manifestation of effects described in this paper (perhaps, in other fields, e.g., in plasma physics, there is a wide range of theory applicability). In the case of relatively big bottom slope, the numerical data provide the results, which are qualitatively similar to those predicted by the asymptotic theory, but quantitatively different (cf. lines 3 and 2 in Fig. 16). When the bottom slope is too big, the incoming solitary wave breaks into a number of secondary solitons (see Fig. 17b); this process is beyond the formal range of applicability of asymptotic theory developed in this paper. Nevertheless, even in this case the dependence of first solitary wave amplitude on distance is close to the prediction of asymptotic theory (cf. lines 4 and 2 in Fig. 16).

The results obtained can be useful for the analysis of observations of internal wave dynamics in coastal zones (see, e.g., [19]). With minor modifications, the ideas and results of this paper can be used in other physical areas, for example, in plasma physics and solid-state physics where similar Ostrovsky equations have been derived.

**Acknowledgements.** The author is thankful to L.A. Ostrovsky for useful discussions. He acknowledges the funding of this study from the State task program in the sphere of scientific activity of the Ministry of Education and Science of the Russian Federation (Project No. 5.1246.2017/4.6) and grant of the President of the Russian Federation for state support of leading scientific schools of the Russian Federation (NSH-2685.2018.5). The author is thankful to the anonymous Referees for the useful remarks and suggestions.

## Appendix

The primitive set of hydrodynamic equations in the long-wave approximation for surface gravity waves is [31]:

$$\frac{\partial \eta}{\partial t} + \nabla_\perp \left[ (h+\eta) \mathbf{q} \right] = 0 \tag{A1}$$

$$\frac{\partial \mathbf{q}}{\partial t} + (\mathbf{q} \cdot \nabla_\perp) \mathbf{q} + [\mathbf{f} \times \mathbf{q}] + g \nabla_\perp \eta = 0, \tag{A2}$$

where $\eta$ is the perturbation of a free surface, $\mathbf{q} = (u, \upsilon)$ is the depth averaged fluid velocity with two horizontal components, longitudinal $u$ and transverse $\upsilon$, $\mathbf{f} = f\mathbf{n}$, where $f = 2\Omega \sin\varphi$ is the Coriolis parameter, $\Omega$ is the angular frequency of Earth rotation, $\varphi$ is the local geographic latitude, $\mathbf{n}$ is the unit vector normal to the Earth surface, and $\nabla_\perp = (\partial/\partial x, \partial/\partial y)$. The similar equations were derived for internal waves [1, 5, 7].

In the linear approximation for waves propagating in the $x$-direction, we obtain from these equations:

$$u = \frac{\omega}{kh}\eta, \quad \upsilon = \frac{if}{\omega}\eta \tag{A3}$$

Then, the wave energy density integrated over a depth $h$ through the cross-section $x =$ constant and averaged over a wave period $T$ is:

$$\langle E \rangle = \frac{1}{T}\int_0^T \left[ \frac{1}{2}\rho \int_{-h}^0 (u^2 + \upsilon^2)\,dz + \rho g \int_0^\eta z\,dz \right] dt = \frac{\rho}{T}\int_0^T \left[ \frac{1}{2}\int_{-h}^0 \left( \left|\frac{\omega}{kh}\eta\right|^2 + \left|\frac{if}{kh}\eta\right|^2 \right) dz + \frac{g}{2}\eta^2 \right] dt$$

$$= \frac{\rho}{2T}\int_0^T \eta^2 dt \left[\left(\left|\frac{\omega}{kh}\right|^2 + \left|\frac{if}{kh}\right|^2\right)h + g\right] = \frac{\rho g}{2T}\int_0^T \eta^2 dt\left(1 + \frac{f^2}{k^2 gh} + 1\right) = \rho g\left(1 + \frac{f^2}{2\omega^2}\right)\langle\eta^2\rangle, \quad (A4)$$

where the angular brackets stand for time averaging. This quantity conserves in the homogeneous fluid, whereas in the spatially inhomogeneous case the conserved quantity is the wave energy flux through the cross-section $x$ = constant, i.e. $J = \langle E\rangle \cdot c(x)$ = constant.

In the case of wave processes described by the Ostrovsky equation (1), it is assumed that $f \ll \omega$, therefore the rotation-induced correction to the wave energy is very small and can be neglected. Then the expression for the wave energy $\langle E\rangle$ coincides with the well-known formula for the non-rotating fluid ($f = 0$) (see, for example, [32, 33] and references therein).

## References

1. L. A. Ostrovsky, Nonlinear internal waves in a rotating ocean, *Oceanology* 18:119–125 (1978).
2. A. I. Leonov, The effect of Earth rotation on the propagation of weak nonlinear surface and internal long oceanic waves, *Ann. New York Acad. Sci*. 373:150–159 (1981).
3. V. M. Galkin and Yu. A. Stepanyants, On the existence of stationary solitary waves in a rotating fluid, *J. Appl. Maths. Mechs*. 55: 939–943 (1991).
4. R.H.J. Grimshaw, J.-M. He, and L.A. Ostrovsky, Terminal damping of a solitary wave due to radiation in rotational systems, *Stud. Appl. Math*. 101:197–210 (1998).
5. R. H. J. Grimshaw, L. A. Ostrovsky, V. I. Shrira, and Yu. A. Stepanyants, Long nonlinear surface and internal gravity waves in a rotating ocean, *Surveys in Geophys*. 19:289–338, (1998).
6. R. Grimshaw, Y. Stepanyants, and A. Alias, Formation of wave packets in the Ostrovsky equation for both normal and anomalous dispersion, *Proc. Roy. Soc., A* 472, 20150416 (2016).
7. K. R. Helfrich, Decay and return of internal solitary waves with rotation, *Phys. Fluids* 19:026601 (2007).
8. R. Grimshaw and K. R. Helfrich, Long-time solutions of the Ostrovsky equation, *Stud. Appl. Math*., 121:71–88 (2008).

9. R. Grimshaw and K. R. Helfrich, The effect of rotation on internal solitary waves, *IMA J. Appl. Math.*, 77:326–339 (2012).

10. R. H. J. Grimshaw, K. R. Helfrich, and E. R. Johnson, Experimental study of the effect of rotation on nonlinear internal waves, *Phys. Fluids* 25:056602 (2013).

11. A. J. Whitfield and E. R. Johnson, Rotation-induced nonlinear wavepackets in internal waves, *Phys. Fluids* 26:056606 (2014).

12. A. J. Whitfield and E. R. Johnson, Wave-packet formation at the zero-dispersion point in the Gardner–Ostrovsky equation, *Phys. Rev. E* 91:051201(R) (2015).

13. O. A. Gilman, R. Grimshaw, and Yu. A. Stepanyants, Approximate analytical and numerical solutions of the stationary Ostrovsky equation, *Stud. Appl. Math.*, 95:115–126 (1995).

14. G. Y. Chen and J. P. Boyd, Analytical and numerical studies of weakly nonlocal solitary waves of the rotation-modified Korteweg–de Vries equation, *Physica D* 155:201–222 (2001).

15. J. P. Boyd and G. Y. Chen, Five regimes of the quasi-cnoidal, steadily translating waves of the rotation-modified Korteweg–de Vries ("Ostrovsky") equation, *Wave Motion* 35:141–155 (2002).

16. L. A. Ostrovsky and Y. A. Stepanyants, Interaction of solitons with long waves in a rotating fluid, *Physica D*, 333:266–275 (2016).

17. O. A. Gilman, R. Grimshaw, and Yu. A. Stepanyants, Dynamics of internal solitary waves in a rotating fluid, *Dynamics. Atmos. and Oceans* 23:403–411 (1996).

18. C. J. Knickerbocker and A. C. Newell, Shelves and the Korteweg–de Vries equation, *J. Fluid Mech.* 98:803–818 (1980).

19. R. Grimshaw, C. Guo, K. Helfrich, and V. Vlasenko, Combined effect of rotation and topography on shoaling oceanic internal solitary waves, *J. Phys. Oceanogr.* 44:1116–1132 (2014).

20. E. Pelinovsky, T. Talipova, I. Didenkulova, and E. Didenkulova (Shurgalina), Long traveling interfacial waves in a two-layer fluid of variable depth, Stud. Appl. Math., this issue (2019).

21. T. G. Talipova, O. E. Kurkina, E. V. Rouvnskaya, and E. N. Pelinovsky, Propagation of solitary internal waves in two-layer ocean of variable depth, *Izvestiya, Atmos. Ocean. Phys.* 51:89–97 (2015).

22. T. G. Talipova, E. N. Pelinovsky, O. E. Kurkina, I. I. Didenkulova, A. A. Rodin, F. S. Pankratov, A. A. Naumov, A. R. Giniyatullin, and I. F. Nikolkina, Propagation of a finite-amplitude wave in a stratified fluid of a variable depth, in Modern Science, *Collection of Papers* (in Russian) 2:144–150 (2012).

23. R. Grimshaw, E. Pelinovsky, and T. Talipova, Solitary wave transformation due to a change in polarity, *Stud. Appl. Math*. 101:357–388 (1998).

24. R. Grimshaw, E. Pelinovsky, and T. Talipova, Solitary wave transformation in a medium with sign-variable quadratic nonlinearity and cubic nonlinearity, *Physica D* 132, 40–62 (1999).

25. L. A. Ostrovsky and J. Grue, Evolution equations for strongly nonlinear internal waves, *Phys. Fluids* 15: 2934–2948 (2003).

26. J. Apel, L. A. Ostrovsky, Y. A. Stepanyants, and J. F. Lynch, Internal solitons in the ocean and their effect on underwater sound, *J. Acoust. Soc. Am.* 121:695–722 (2007).

27. L. A. Ostrovsky, E. N. Pelinovsky, V. I. Shrira, and Y. A. Stepanyants, Beyond the KDV: Post-explosion development, *Chaos*, 25: 097620 (2015).

28. M. Obregon and Y. Stepanyants, On numerical solution of the Gardner–Ostrovsky equation, *Math. Mod. Nat. Proc*. 7:113–130 (2012).

29. L. A. Ostrovsky and Yu. A. Stepanyants, Nonlinear surface and internal waves in rotating fluids. In *Nonlinear Waves* 3. *Physics and Astrophysics*, Proc. 1989 Gorky School on Nonlinear Waves, eds A. V. Gaponov-Grekhov, M. I. Rabinovich and J. Engelbrecht. Springer-Verlag, Berlin–Heidelberg (1990) 106–128.

30. M. Obregon, N. Raj N., and Y. Stepanyants, Dynamics of Gardner solitons under the influence of the Earth' rotation, *Chaos*, 28: 033106 (11 p.) (2018).

31. V. I. Shrira, Propagation of long nonlinear waves in a layer of rotating fluid, *Izvestiya. Atmos. Ocean. Phys*., 17: 55–59 (1981).

32. P. Maïssa, G. Rousseaux, and Y. Stepanyants, Negative energy waves in shear flow with a linear profile. *Eur. J. Mech. – B/Fluids*, 56: 192–199 (2016).

33. S. Churilov, A. Ermakov, G. Rousseaux, and Y. Stepanyants, Scattering of long water waves in a canal with rapidly varying cross-section in the presence of a current, *Phys. Rev. Fluids*, 2:094805, 18 p. (2017).